\documentclass[%
prl,
 reprint,
 amsmath,amssymb,amsfonts
 aps,preprintnumbers]{revtex4-1}
\pdfoutput=1

\usepackage[english]{babel}
\usepackage{epsf,epsfig}
\usepackage{amssymb,amsmath,amsfonts}
\usepackage{mathrsfs} 
\usepackage{enumitem}

\usepackage{tikz}

\usepackage[mathscr]{eucal}

\usetikzlibrary{decorations.pathreplacing}

\def\cO {\mathcal{O}}
\def\cP {\mathcal{P}}

\def\cB{\mathcal{B}}
\def\cA{\mathcal{A}}
\def\cM{\mathcal{M}}
\def\cF{\mathcal{F}}
\def\cI{\mathcal{I}}
\def\cH{\mathcal{H}}

\def\correlator{\langle p_1 p_2 p_3 p_4\rangle}

\def\xb{\bar{x}}
\def\yb{\bar{y}}

\def\tw{s}
\def\CB{\mathcal{B}}
\def\bF{\mathbf{F}}
\def\bP{\mathbf{P}}
\def\LL{\mathbb{L}}

\def\var{s}
\def\z{z}
\def\w{w}
\def\q{h}

\newcommand{\beq}{\begin{equation}}
\newcommand{\eeq}{\end{equation}}
\newcommand{\bea}{\begin{eqnarray}}
\newcommand{\eea}{\end{eqnarray}}

\newcommand{\nn}{\nonumber}

\begin{document}

\title{The double-trace spectrum of $\mathcal{N}=4$ SYM at strong coupling}

\author{Francesco Aprile${}^1$,  James Drummond${}^2$, Paul Heslop${}^3$,  Hynek Paul${}^2$}
\affiliation{
$\rule{0pt}{.5cm}$
${}^{1}$Dipartimento di Fisica, Universit\`a di Milano-Bicocca \& INFN, Sezione di Milano-Bicocca, I-20126 Milano,\\
${}^{2}$School of Physics and Astronomy,  University of Southampton, Highfield,  SO17 1BJ,\\
${}^{3}$Mathematics Department, Durham University, Science Laboratories, South Rd, Durham DH1 3LE}

\begin{abstract}
\noindent 
The spectrum of IIB supergravity on AdS${}_5 \times S^5$ contains a number of bound states described 
by long double-trace multiplets in $\mathcal{N}=4$ super Yang-Mills theory at large 't Hooft coupling. 
At large $N$ these states are degenerate and to obtain their anomalous dimensions as expansions 
in $\tfrac{1}{N^2}$ one has to solve a mixing problem.  
We conjecture a formula for the 
leading anomalous dimensions of all long double-trace operators which exhibits a large residual degeneracy whose structure we describe. 
Our formula can be related to conformal Casimir operators which arise in the structure of leading discontinuities of supergravity loop corrections to four-point correlators of half-BPS operators.
\end{abstract}
\maketitle

%%%%%%%%%%%%%%%%%%%%%%%%%%%%%%%%%%%%%%%%%%%%%%%%%%%%%%%
\section{I.\ Introduction}
%%%%%%%%%%%%%%%%%%%%%%%%%%%%%%%%%%%%%%%%%%%%%%%%%%%%%%%

Recently much progress has been made in understanding the structure of the spectrum 
of double-trace operators in $\mathcal{N}=4$ super Yang-Mills theory at large $N$ and 
large 't Hooft coupling $\lambda=g^2N$ \cite{Aprile:2017xsp}. 
Based on these results, OPE and bootstrap techniques have been applied in \cite{Aprile:2017qoy,Aprile:2017bgs} 
to obtain closed form expressions for supergravity loop corrections of certain holographic correlators, uncovering novel and rich structure (see \cite{Alday:2017xua,Alday:2017vkk} for related approaches to such loop corrections).
Here we complete the picture for the double-trace spectrum and conjecture a general formula for the leading anomalous dimensions of all long double-trace operators of any twist, spin and $su(4)$ representation.

In the regime $N\rightarrow \infty$ and $\lambda\gg 1$, the theory is in correspondence with classical IIB supergravity on ${\rm AdS}_5\times S^5$ \cite{Maldacena:1997re}.
The graviton and the Kaluza-Klein multiplets are dual to protected half-BPS operators in the $[0,p,0]$ representation of $su(4)$,
\beq
\label{Opdef}
\cO_p=y^{i_1}\ldots y^{i_p}\mathrm{Tr}\left( \Phi_{i_1}\ldots\Phi_{i_p} \right) + \ldots
\eeq
where $\Phi_{i=1,\ldots 6}$ are the elementary scalar fields, the complex vector $\vec{y}\in SU(4)/ S(U(2)\times U(2))$, 
and the ellipsis stands for $1/N$-suppressed multi-trace terms (for $p \geq 4$), 
whose precise nature will be described in Section~II.

At leading large $N$ (for any value of $\lambda$)
we may consider degenerate long double-trace superconformal primary operators of twist $\tau$, spin $l$ and $su(4)$ labels $[a,b,a]$ of the form
\begin{equation}
\mathcal{O}_{pq} = \mathcal{O}_{p} \partial^l \Box^{\frac12(\tau-p-q)}  \mathcal{O}_{q} \,, \qquad (p \leq q)\,.
\label{ops}
\end{equation} 
The $d$ allowed values of the pair $(p,q)$ run over a set $\mathcal{D}^{\rm long}_{\tau,l,a,b}$. We parametrise this set by $i,r$ as follows:
\begin{align}\label{ir}
	p&=i+a+1+r\,, \qquad &q&=i+a+1+b-r\,, \notag\\
	i&=1,\ldots,(t-1)\,, &r&=0,\ldots,(\mu -1)\,,
\end{align}
so that $d=\mu(t-1)$ with
\beq\label{multiplicity}
t\equiv (\tau-b)/2-a\,,\quad 
\mu \equiv   \left\{\begin{array}{ll}
\bigl\lfloor{\frac{b+2}2}\bigr\rfloor \quad &a+l \text{ even,}\\[.2cm]
\bigl\lfloor{\frac{b+1}2}\bigr\rfloor \quad &a+l \text{ odd.}
\end{array}\right.
\eeq

The operators $\mathcal{O}_{pq}$ are in long multiplets, but in the strict large $N$ limit their dimensions are protected. 
At order $1/N^2$ they acquire anomalous dimensions and mix amongst themselves and with other long operators. 
In the supergravity regime $\lambda \gg 1$, operators  corresponding to massive string excitations 
should decouple from the spectrum leaving only those corresponding to supergravity states, e.g. the single-particle states $\mathcal{O}_p$ and the two-particle bound states $\cO_{pq}$. 
At leading order in large $N$ the $\mathcal{O}_{pq}$ just mix amongst themselves 
to produce the true scaling eigenstates, which we denote by $K_{pq}$. 
Mixing with higher multi-particle states will only occur at higher orders in the $1/N$ expansion.
Analysis of the OPE in the tree-level supergravity regime (see  Section~III) leads us to the following conjecture, 
generalising results in \cite{Aprile:2017bgs,Aprile:2017xsp,Aprile:2017qoy}.

{\bf Main conjecture.} 
Up to order $1/N^2$, the dimensions of the operators $K_{pq}$ are given by
\beq\label{anomdims}
\begin{minipage}{0.9\linewidth}
\centering
\begin{tikzpicture}
\def\ox{0}
\def\oy{0}
\def\passo{1}
\draw (\ox,\oy) node [draw]  {
$
\displaystyle
\Delta_{pq}=\tau + l - \frac{2}{N^2} \frac{2 M^{(4)}_{t} M^{(4)}_{t+l+1}}{ \left(l+2p-2-a - \frac{1+(-)^{a+l}}{2} \right)_6}
$
\rule{.2cm}{0pt}
};
\end{tikzpicture}
\end{minipage}
\eeq
Here $(\ldots)_6$ is the Pochhammer symbol, and we define
\beq
M^{(4)}_t\equiv (t-1) (t+a) (t+a+b+1)(t+2a+b+2) \,.
\eeq
Note that for $\mu>1$ and $t> 2$ some dimensions exhibit a residual degeneracy because they are independent of $q$.
We display this property with an illustration of $\mathcal{D}^{\rm long}_{\tau,l,a,b}$. 
%
% FIG WITH LEGEND
% 
\beq
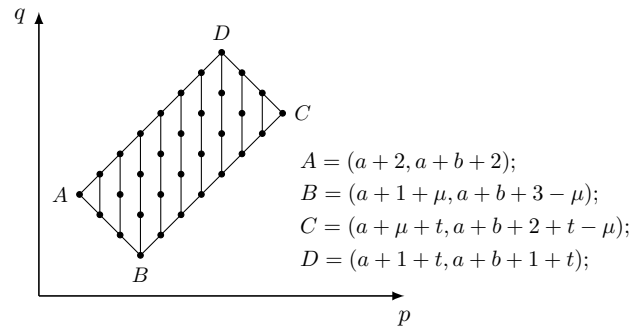
\begin{tikzpicture}[scale=.54]
\def\prop{.5}
\def\shifthor{\prop*2}
\def\oxy{(\prop*2.-\shifthor-1,\prop*5-1)};
\def\oxxy{(\prop*18-\shifthor,\prop*5-1)}
\def\oxyy{(\prop*2.-\shifthor-1,\prop*17)}
\def\ptuno{(\prop*2-\shifthor,\prop*8)}
\def\ptdue{(\prop*5-\shifthor,\prop*5)}
\def\pttree{(\prop*9-\shifthor,\prop*15)}
\def\ptquattro{(\prop*12-\shifthor,\prop*12)}
%
%axis horizontal
\draw[-latex, line width=.6pt]		\oxy    --   \oxxy ;
\node[scale=.8] (oxxy) at (\prop*18-\shifthor,\prop*5.25) {};
\node[scale=.9] [below of=oxxy] {$p$};
%
%axis vertical
\draw[-latex, line width=.6pt] 		\oxy    --   \oxyy;
\node[scale=.8] (oxyy) at (\prop*4-\shifthor*1.8,\prop*16.8) {};
\node[scale=.9] [left of= oxyy] {$q$};
%
%rectangle
\draw[] 						\ptuno -- \ptdue;
\draw[]						\ptuno --\pttree;
\draw[]						\ptdue --\ptquattro;
\draw[]						\pttree--\ptquattro;
%
%dots
%
\foreach \indeyc in {0,1,2,3}
\foreach \indexc  in {2,...,9}
\filldraw   					 (\prop*\indexc+\prop*\indeyc-\shifthor, \prop*6+\prop*\indexc-\prop*\indeyc)   	circle (.07);
%
%letters
%
\node[scale=.8] (puntouno) at (\prop*4-\shifthor,\prop*8) {};
\node[scale=.8]  [left of=puntouno] {$A$};   
\node[scale=.8] (puntodue) at (\prop*5-\shifthor,\prop*6+.5) {};
\node[scale=.8] [below of=puntodue]  {$B$}; 
\node[scale=.8] (puntoquattro) at (\prop*13-\shifthor,\prop*15) {};
\node[scale=.8] [below of=puntoquattro] {$C$};
\node[scale=.8] (puntotre) at (\prop*9-\shifthor,\prop*13) {};
\node[scale=.8] [above of=puntotre] {$D$}; 
%
%
%legend
\node[scale=.84] (legend) at (\prop*19,\prop*7) {$\begin{array}{l}  
													\displaystyle A=(a+2,a+b+2); \\[.1cm]
													\displaystyle B=(a+1+\mu,a+b+3-\mu); \\[.1cm]
													\displaystyle C=(a+\mu+t, a+b+2+t-\mu); \\[.1cm]
													\displaystyle D=(a+1+t,a+b+1+t); \\[.1 cm] \end{array}$  };
%									
%lines
%
\foreach \indexc in {3,4}
\draw (\prop*\indexc-\shifthor, \prop*10-\prop*\indexc ) -- (\prop*\indexc-\shifthor, \prop*6+\prop*\indexc  );
\foreach \indexc in {5,6,7,8,9}
\draw (\prop*\indexc-\shifthor, \prop*\indexc ) -- (\prop*\indexc-\shifthor, \prop*6+\prop*\indexc  );
\foreach \indexc in {10,11,12}
\draw (\prop*\indexc-\shifthor, \prop*\indexc ) -- (\prop*\indexc-\shifthor, \prop*24-\prop*\indexc  );
\end{tikzpicture}
\nn
\eeq
The dots connected by vertical lines represent operators of common anomalous dimension.
%
%

%
%%%%%%%%%%%%%%%%%%%%%%%%%%%%%%%%%%%%%%%%%%%%%%%%%%%%%%%
\section{II. Holographic correlators}
%%%%%%%%%%%%%%%%%%%%%%%%%%%%%%%%%%%%%%%%%%%%%%%%%%%%%%%
%
%
The correlators $\langle \cO_{p_1}\cO_{p_2}\cO_{p_3}\cO_{p_4} \rangle\equiv \langle p_1 p_2 p_3 p_4\rangle $ may be written as a free part plus an interacting part,
\bea
\label{corinfs}
\correlator 									&=& \correlator _{\rm free} + \cP \times \cI \times \cH \,.
\eea
The factor $\mathcal{P}$ carries the conformal and $su(4)$ weights and assuming (without loss of generality) $p_{21} \ge 0$, $p_{43} \ge 0$ and $p_{43}\ge p_{21}$, it takes the form
\beq
\cP= N^{\frac{1}{2}\sum p_i } g_{12}^{\frac{p_1+p_2-p_{43} }{2} } \,g_{14}^{\frac{-p_{21}+p_{43}}{2}}\,  g_{24}^{\frac{p_{21}+p_{43}}{2} } g_{34}^{{p_3}{} }\, ,
\eeq
where $p_{ij}=p_i-p_j$ and $g_{ij}= {(y_i \cdot y_j) }\big/{ {x}_{ij}^{\,2} }$.
The quantities $\cI$ and $\cH$ are functions of the variables $x,\bar{x},y,\bar{y}$, related to the conformal and $su(4)$ cross-ratios $u,v,\sigma,\tau$ via
\begin{alignat}{3}
u=x \bar{x} &= \frac{ {x}^{\,2}_{12} {x}^{\,2}_{34} }{  {x}^{\,2}_{13} {x}^{\,2}_{24} } \,, \quad v=&&(1-x)(1-\bar{x})&&=\frac{ {x}^{\,2}_{14} {x}^{\,2}_{23} }{{x}^{\,2}_{13} {x}^{\,2}_{24} }, \notag \\
\frac{1}{\sigma}=y \bar{y} &= \frac{ {y}^{\,2}_{12} {y}^{\,2}_{34} }{  {y}^{\,2}_{13} {y}^{\,2}_{24} } \,, \quad \frac{\tau}{\sigma}=&&(1-y)(1-\bar{y})&&=\frac{ {y}^{\,2}_{14} {y}^{\,2}_{23} }{{y}^{\,2}_{13} {y}^{\,2}_{24} }\,.
\end{alignat}
In terms of these variables we have
\begin{equation}
\cI(x,\xb,y,\yb)=(x-y)(x-\yb)(\xb-y)(\xb-\yb){ \Big/ (y \yb)^2}.
\end{equation}

The decomposition into free and interacting parts in (\ref{corinfs}) reflects the property of `partial non-renormalisation' \cite{Eden:2000bk}, i.e. 
the statement that all the dependence on the coupling appears in the function $\cH$. Here we consider the leading contribution to $\cH$ at large $\lambda$.
In the OPE of $(\mathcal{O}_{p_1}  \times \mathcal{O}_{p_2})$ and $(\mathcal{O}_{p_3}  \times \mathcal{O}_{p_4})$,
 the free term contributes both a protected sector, and a long sector.  Identifying the sectors
is non-trivial due to possible semishort multiplet recombination at the unitarity bound \cite{Dolan:2002zh,Doobary:2015gia}.

At leading order in the $1/N^2$ expansion, a correlator is determined by disconnected contributions to the free part. 
These only exist for $\langle ppqq \rangle$ and cases related by crossing,
\bea
\label{discoprop}
 \langle p p q q\rangle_{} = pq \mathcal{P} \! \left[ 1+ \delta_{pq} \Bigl[ \Bigl(\frac{g_{13} g_{24}}{g_{12} g_{34}}\Bigr)^p \! + \!  \Bigl( \frac{ g_{14} g_{23} }{ g_{12} g_{34} } \Bigr)^p \Bigr] \right] . \quad 
\eea
At the next order in $1/N^2$ in the supergravity regime, tree-level Witten diagrams contribute both the free theory connected diagrams and the first contribution to $\mathcal{H}$.

{\bf Supergravity states and free theory.}
It was noticed in \cite{Uruchurtu:2008kp} that the connected part of $\langle p_1 p_2 p_3 p_4 \rangle_{\rm free}$ generated via tree-level Witten diagrams,
disagrees with free theory four-point functions of single trace half-BPS operators. The resolution is that {\it single-particle supergravity 
states are not dual to single trace half-BPS operators, rather they are uniquely defined as those orthogonal to all multi-trace operators}.
From this property we can identify multi-trace contributions to ${\rm Tr}\,\Phi^p$ for $p \geq 4$. The presence of multi-trace admixtures was also discussed in \cite{Arutyunov:1999en,Rastelli:2017udc}.
Consider
for example $\cO_4$, the condition $\langle \mathcal{O}_4 (\mathcal{O}_2)^2\rangle= 0$ determines, 
\begin{equation}
\label{multi-traces}
\mathcal{O}_4 = y^{i_1}\ldots y^{i_4}\mathrm{Tr}\left( \Phi_{i_1}\ldots\Phi_{i_4} \right) - \frac{2N^2-3}{N(N^2+1)}(\mathcal{O}_2)^2\,.
\end{equation}
With this identification of $\cO_4$ the free theory computation of $\langle 2244\rangle$ agrees with that of supergravity \cite{Uruchurtu:2008kp}. 
The correct identification of the operators $\mathcal{O}_p$ is also necessary for the `derivative relation' of \cite{0907.0151} to hold, as can be directly observed for the cases $\langle 22nn \rangle$.

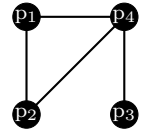
\begin{figure}
\begin{tikzpicture}[scale=.65]
\def\x1{0}
\def\y1{1}
\def\colore{black}

\draw[\colore, thick] (\x1,\y1)--(\x1+2,\y1+2);
\draw[\colore, thick] (\x1,\y1)--(\x1,\y1+2);
\draw[\colore, thick] (\x1,\y1+2)--(\x1+2,\y1+2);
\draw[\colore, thick] (\x1+2,\y1+2)--(\x1+2,\y1);

\filldraw (\x1,\y1) 		circle (.28);
\filldraw (\x1+2,\y1) 		circle (.28);
\filldraw (\x1,\y1+2) 		circle (.28);
\filldraw (\x1+2,\y1+2)	circle (.28);

\draw[text=white, text centered, font=\footnotesize] (\x1,\y1)         	node     { p$_2$};
\draw[text=white, text centered, font=\footnotesize] (\x1+2,\y1)     	node     { p$_3$};
\draw[text=white, text centered, font=\footnotesize] (\x1,\y1+2)     	node     { p$_1$};
\draw[text=white, text centered, font=\footnotesize] (\x1+2,\y1+2) 	node     { p$_4$};

\end{tikzpicture}
\caption{A free theory diagram absent from $\langle p_1 p_2 p_3 p_4\rangle$.}
\label{vantop}
\end{figure}

More generally,  connected free theory diagrams where e.g. $O_{p_3}$ is joined only to $\mathcal{O}_{p_4}$ (see Fig. \ref{vantop}) are absent.
To see this note that at twist $p_{43}$ in the $(\cO_{p_3} \times  \cO_{p_4})$ OPE, only a half-BPS operator $\mathcal{O}_{p_{43}}$ of charge $p_{43}$ could potentially be transferred.
By our definition, $\cO_{p_4}$ is orthogonal to all multi-trace operators and in particular
to the double (or higher) trace operator $[\cO_{p_{43}} \cO_{p_3}]$.
But the vanishing two-point function $\langle [\cO_{p_{43}} \cO_{p_3}] \cO_{p_{4}} \rangle$ is just a non-singular limit of the three-point function, 
$\langle \cO_{p_{43}}  \cO_{p_3} \cO_{p_{4}} \rangle$, which therefore  also vanishes. Hence no operator $\mathcal{O}_{p_{43}}$ 
can be exchanged and the coefficient of the above diagram must vanish. Note that this holds no matter if $\cO_{p_{43}}$ is single-trace, 
multi-trace or a combination thereof. Obviously any topology related by a permutation to Fig. \ref{vantop} also vanishes.

{\bf Tree level dynamics.} 
The conjecture of \cite{Rastelli:2016nze} is a simple Mellin integral for the leading term in $\mathcal{H}$: 
\bea
\label{Rastelli_corr}
\!\!\!\mathcal{H}_{\rm RZ}&=& -\mathcal{N}_{p_1p_2p_3p_4} \oint d\z d\w\, u^{\frac{\z}{2} } v^{\frac{\w}{2}} \mathcal{R}[\substack{\z\ \w\\ \sigma\ \tau}]\, \Gamma_{p_1p_2p_3p_4}, \notag \\
\Gamma&=&\Gamma[\tfrac{p_1+p_2-\z}{2}]\Gamma[\tfrac{p_3+p_4-\z}{2}]\Gamma[\tfrac{p_1+p_4-\w}{2}]\times \notag \\
& & \Gamma[\tfrac{p_2+p_3-\w}{2}]\Gamma[\tfrac{\z+\w+4-p_1-p_3}{2}]\Gamma[\tfrac{\z+\w+4-p_2-p_4}{2}],\notag\\
\mathcal{R}&=&\frac{ u^{ \frac{ p_3-p_4}{2} } }{ v^{\frac{ p_2+p_3}{2} } } \! \sum_{i,j} \! \frac{ a_{ijk} }{ i!j!k!} \frac{ \sigma^i \tau^j  (\tilde{\mu}-\z-\w+2i)^{-1}}{(\z-\tilde{\z}+2k)(\w-\tilde{\w}+2j)}.\,\,\,\,\,\,\,
\eea
In the sum $i,j,k\geq0$ and we use the notation:
\begin{alignat}{3}
\tilde{\mu}&=p_2 +  p_4 - 2, &&\tilde{\w}&&=p_2+p_3-2,\notag \\
\tilde{\z}&={\rm min}(p_1 + p_2,p_3 + p_4)-2,\,\,\,\,\,\, &&k&&=M-1-i-j,\notag \\
\!\!\!\!\!\! M&=p_3-1+{\rm min}(0,\Lambda), &&\Lambda &&= \tfrac{p_1+p_2-p_3-p_4}{2}. 
\end{alignat}
Finally, the coefficients $a_{ijk}$ are given by
\beq
\label{aijkRastelli}
a_{ijk}=
\frac{ 2^3 (M-1)! }{(1+|\Lambda |)_k (1+\frac{p_{43}+p_{21}}{2})_i (1+\frac{p_{43}-p_{21}}{2})_j }\ .
\eeq
The conjecture agrees with all known supergravity computations  (\cite{Arutyunov:2017dti} and refs. therein).
The assumptions which led to \eqref{Rastelli_corr} are spelled out in \cite{Rastelli:2017udc}.

{ \bf Determining $\mathcal{N}_{p_1p_2p_3p_4}$ from the light-like limit.}
The normalisation $\mathcal{N}$ is not determined in \cite{Rastelli:2016nze}.
Here we fix it using the following non-trivial statement: 
\beq\label{lightlikelimit}
\begin{minipage}{0.8\linewidth}
\centering
\begin{tikzpicture}
\def\ox{0}
\def\oy{0}
\def\passo{1}
\draw (\ox,\oy) node [draw]  {
\rule{.2cm}{0pt}
$\displaystyle
\lim_{ \substack{ u,v\rightarrow 0} } \frac{\correlator}{\mathcal{P}}\Big|_{\textstyle \frac{1}{N^2}}=0$, \rule{0.2cm}{0pt} $\displaystyle \frac{u}{v}$ fixed.
\rule{.2cm}{0pt}};
\end{tikzpicture}
\end{minipage}
\eeq
The limit $u,v\rightarrow 0$ with $(u/v)$ fixed corresponds to taking the points $x_1,x_2,x_3,x_4$ to be sequentially light-like separated.

Examining both the free theory and interacting contributions to the LHS of (\ref{lightlikelimit}) above, we find that it takes the form $\sum_{r=1}^{M} A_r (u \tau/v)^r$ where 
\begin{align} \label{A}
A_{r}= p_1p_2p_3p_4 \frac{p_{21}+p_{43} +2}{2 N^2} - \mathcal{N}_{p_1p_2p_3p_4}R_{p_1p_2}^{p_3p_4}\,.
\end{align}

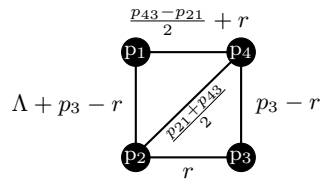
\begin{figure}
%\begin{array}{c}
\begin{tikzpicture}[scale=.7]
\def\x1{0}
\def\y1{1}
\def\colore{black}

%%%%%%%%%%%%%%%%%%%%%%%%%%%%%
\draw[\colore, thick] (\x1,\y1+.05)--(\x1+2,\y1+2);
\node[rotate=45] at (\x1+1.2,\y1+.8) 					{$\frac{p_{21}+p_{43}}{2}$ };

\draw[\colore, thick] (\x1,\y1)--(\x1,\y1+2);
\node[rotate=0] at (\x1-1.3,\y1+1)  					{ $\Lambda+p_3-r$}; 

\draw[\colore, thick] (\x1,\y1+2)--(\x1+2,\y1+2);
\node[]  at (\x1+1,\y1+2.6) 						{ $\tfrac{p_{43}-p_{21}}{2}+r$ };  

\draw[\colore, thick] (\x1+2,\y1+2)--(\x1+2,\y1);
\node[rotate=0] at (\x1+2.9,\y1+1)					 {  ${p_3-r}{} $  };

\draw[\colore, thick] (\x1,\y1)--(\x1+2,\y1);
\node[] at (\x1+1,\y1-.3) 							{$r$};

\filldraw (\x1,\y1) 		circle (.27);
\filldraw (\x1+2,\y1) 		circle (.27);
\filldraw (\x1,\y1+2) 		circle (.27);
\filldraw (\x1+2,\y1+2)	circle (.27);
\draw[text=white, text centered, font=\footnotesize] (\x1,\y1)         	node     { p$_2$};
\draw[text=white, text centered, font=\footnotesize] (\x1+2,\y1)     	node     { p$_3$};
\draw[text=white, text centered, font=\footnotesize] (\x1,\y1+2)     	node     { p$_1$};
\draw[text=white, text centered, font=\footnotesize] (\x1+2,\y1+2) 	node     { p$_4$};
\node[font=\footnotesize] at (\x1+4,\y1-.25) {$\phantom{space}$};
\end{tikzpicture}
\caption{Free theory diagrams in the light-like limit.}
\label{diagrams1}
\end{figure}

The first term in (\ref{A}) comes from $\langle p_1 p_2 p_3 p_4 \rangle_{\rm free}/\mathcal{P}$ and arises from the diagrams in Fig. \ref{diagrams1}. 
{The normalisation of each of these diagrams in the planar limit can be simply obtained by counting the number of inequivalent planar embeddings.}
Cyclic rotation on each vertex leaves the diagram unchanged, hence the factor $p_1p_2p_3p_4$. 
Additionally, the diagonal propagators can be drawn inside or outside the square, giving $\tfrac{1}{2}(p_{21} +p_{43})+1$ different possibilities. 
The multi-trace terms in $\mathcal{O}_p$ do not affect the leading $N$ result for the diagram. 
The cases $r=0$ or $r=M+1$ correspond to the diagrams of Fig. \ref{vantop} which are absent as discussed above.

The second contribution in (\ref{A}) is obtained from $\mathcal{I}\times\mathcal{H}_{\rm RZ }$.
Note that each term in $u^{\frac{\z}{2} } v^{\frac{\w}{2}} \mathcal{R}$
has the form
\beq
\frac{  u^{\frac{\z-p_{43}}{2} } v^{\frac{\w-p_2-p_3}{2}} \sigma^i \tau^j  }{( \z-p_{43} -2-2(i+j) )(\w-p_2-p_3+2+2j)},
\eeq
and upon residue integration will produce a term proportional to $(u\sigma)^i (u/v)^{1+j}\tau^j$.
Since $\mathcal{I}=\tau+O(u,v)$, we find that the contribution to $A_{r}$ comes from taking the simple poles with $i=0$ in \eqref{aijkRastelli}. 
The residue is 
\bea
R_{p_1p_2}^{p_3p_4}&=& |\Lambda |! (\tfrac{p_{43}-p_{21}}{2})!( \tfrac{p_{21}+p_{43}+2}{2} )!  ( M-1)!\,.
\eea
Crucially the $j$ dependence cancels between $a_{0jk}/(j!k!)$ and $\Gamma_{p_1p_2p_3p_4}$ and hence $A_r$ is in fact independent of $r$. 
Now the statement \eqref{lightlikelimit} is clearly equivalent to the statement $A_{r}=0$ for all $r$.
Rearranging \eqref{A} we thus obtain the result for $\mathcal{N}_{p_1 p_2 p_3 p_4}$,
\beq
\label{N}
\mathcal{N} =
\frac{1}{N^2} \frac{p_1 p_2 p_3 p_4 }{ | \Lambda |! ( \tfrac{p_{43}-p_{21}}{2})!( \tfrac{p_{43}+p_{21}}{2} )! (M-1)!}\,.
\eeq
The result combines neatly with the coefficients $a_{ijk}$,
\beq
\mathcal{N} a_{ijk} =\! \frac{1}{N^2}  \frac{2^3 p_1 p_2 p_3 p_4}{(|\Lambda|+k)!(\frac{p_{43}+p_{21}}{2}+i)! (\frac{p_{43}-p_{21}}{2}+j)!}.
\eeq
Note that the expression (\ref{N}) is consistent with the results for $\mathcal{N}_{ppqq}$ and $\mathcal{N}_{p,p+1,q,q+1}$ obtained in \cite{Aprile:2017xsp,Aprile:2017qoy}.

{\bf Proof of light-like vanishing.} 
The light-like limit projects the common OPE of $(\mathcal{O}_{p_1}  \times \mathcal{O}_{p_2})$ and $(\mathcal{O}_{p_3}  \times \mathcal{O}_{p_4})$ onto operators with large spin and twist $\tau \leq p_{43}+2M$, i.e. twist $\tau < {\rm min}(p_1+p_2,p_3+p_4)$.
To justify the statement (\ref{lightlikelimit}) let us consider the various contributions to the OPE expected in the supergravity regime. First of all we have single-particle states corresponding to half-BPS superconformal primary operators. Such operators have spin zero and do not contribute in the limit $v \rightarrow 0$ which receives contributions from large spin. Next we have (both protected and unprotected) double-trace operators of the form $[\mathcal{O}_p \Box^n \partial^l \mathcal{O}_q]$ or mixtures thereof. The leading large $N$ contribution to three-point functions of the form $\langle \mathcal{O}_p \mathcal{O}_q [\mathcal{O}_{p'} \Box^n \partial^l \mathcal{O}_{q'}]\rangle \sim O(N^{p+q})$ arises when $p=p'$ and $q=q'$ when the three point function factorises into a product of two-point functions. The twist $\tau$ of the double-trace operator therefore must obey $\tau \geq p+q$, otherwise the three-point function will be suppressed by $1/N^2$. 
The exchanged operators surviving the light-like limit (\ref{lightlikelimit}) all have twist less than both $p_1+p_2$ and $p_3+p_4$ and hence the contributions will be suppressed by at least $1/N^4$ and will not contribute to the LHS of (\ref{lightlikelimit}). 
Higher multi-trace operators are even more suppressed and we conclude that no operators in the supergravity spectrum can contribute in the light-like limit, justifying (\ref{lightlikelimit}).

%%%%%%%%%%%%%%%%%%%%%%%%%%%%%%%%%%%%%%%%%%%%%%%%%%
\section{III. Unmixing Equations}
 %%%%%%%%%%%%%%%%%%%%%%%%%%%%%%%%%%%%%%%%%%%%%%%%%%

We now describe how the system of relations implied by the OPE describes an eigenvalue problem which allows us to determine the anomalous dimensions of the true double-trace eigenstates $K_{pq}$.
In particular, we consider the long multiplet SCPW expansion
of the correlators $\correlator$, in which the pairs $(p_1,p_2)$ and $(p_3,p_4)$ both run over the set $\mathcal{D}_{\tau,l,a,b}^{\rm long}$ described in (\ref{ir}). 
The result is a symmetric $(d\times d)$ matrix whose partial wave expansion reads
\beq
\label{longscpwexp}
\Big[ { \langle p_1 p_2 p_3 p_4\rangle}\Big]=\sum_{\tau,l,a,b}\Big[\mathbb{\cA}^{\tau,l}_{a,b} + 
												\tfrac{1}{N^2} \! \log u \, \mathbb{\cM}^{\tau,l}_{a,b} \Big]  \LL_{[a,b,a]}^{(\tau|l)}  .
\eeq
Terms of order $1/N^2$ which are analytic at $u=0$, i.e. without a factor of $\log u$, have been dropped on the RHS. 

The matrix $\mathbb{\cA}^{\tau,l}_{a,b}$ in (\ref{longscpwexp}) is determined by disconnected free theory and is diagonal due to the form  of the disconnected contributions (\ref{discoprop}). The matrix
$\mathbb{\cM}^{\tau,l}_{a,b}$ is obtained from the discontinuity around $u=0$ of $\cH_{RZ}$.
For completeness, we recall the explicit expression \cite{Dolan:2001tt,Arutyunov:2002fh} of a
long supermultiplet of twist $\tau$, spin $l$ and $su(4)$  rep $\mathfrak{R}=[n-m,2m+p_{43},n-m]$, 
\beq\label{LONGBL}
\LL^{(\tau|l)}_{\mathfrak{R}} =\mathcal{P}\,\cI(x,\bar{x},y,\bar{y}) \frac{  {\Upsilon}_{nm}(y,\bar{y})\, \cB^{\, 2+\frac{\tau}{2} |l }(x,\xb)   }{ u^{2+\frac{p_{43} }{2} } }\ .
\eeq
This structure is the simplest among the determinantal superconformal blocks \cite{Doobary:2015gia}, 
since it factorises into an ordinary conformal block $\CB^{\,\tw|l}(x,\xb)$ \cite{Dolan:2000ut},  
\bea
\label{Confblock}
\CB^{\,\tw|l}(x,\xb)&=&(-)^l\  \frac{u^\tw x^{l+1}\, \bF_{\tw+l }(x)\bF_{\tw-1}(\xb)- (x\leftrightarrow \xb)  }{x-\xb}, \nn\\ 
\bF_{\tw}(x)&=&~_2F_1\big[{\tw- \tfrac{p_{12}}{2}, \tw+\tfrac{p_{34}}{2}; 2\tw}\big](x),
\eea
and an $su(4)$ block $\Upsilon_{nm}(y,\yb)$ \cite{Nirschl:2004pa}, 			 
\bea				 
\label{SU4harm}
\Upsilon_{nm}(y,\yb)&=& - \frac{ \bP_{n+1}(y) \bP_m(\yb)-  \bP_m(y) \bP_{n+1}(\yb)  }{y-\yb},\\
\bP_{n}(y)&=&\frac{n!\, y}{(n+1+p_{43})_n}\, {\rm JP}^{(p_{43}-p_{21}|p_{43}+p_{21})}_n\left(\tfrac{2}{y}-1 \right), \nn
\eea
where ${\rm JP}$ stands for a Jacobi polynomial.

The matrices $\mathcal{A}$ and $\mathcal{M}$ contain CFT data for the operators $K_{pq}$:
\begin{align}
\label{eigen_probl}
\mathbb{\cA}^{\tau,l}_{a,b} &= \mathbb{C}_{\tau,l,a,b}\cdot \mathbb{C}^T_{\tau,l,a,b}\,, \notag \\
\mathbb{\cM}^{\tau,l}_{a,b} &= \mathbb{C}_{\tau,l,a,b}\cdot \eta \cdot \mathbb{C}^T_{\tau,l,a,b}\,.
\end{align}
Here the $(d\times d)$ matrix $\mathbb{C}$, indexed by pairs $(p_1,p_2)$ and $(q_1,q_2)$ running over $\mathcal{D}^{\rm long}_{\tau,l,a,b}$, is given by
\beq\label{def3pointfuncmatrix}
\mathbb{C} \equiv \Big[ \langle \mathcal{O}_{p_1} \mathcal{O}_{p_2} K_{q_1q_2}\rangle\Big]\,,
\eeq
and $\eta = {\rm diag}(\eta_{pq})$ is a $(d \times d)$ diagonal matrix where $\eta_{pq}$ is (half) the anomalous dimension of the operator $K_{pq}$ for $(p,q)\in \mathcal{D}^{\rm long}_{\tau,l,a,b}$,
\beq
\Delta_{pq}=\tau+ l + \frac{2}{N^2} \eta_{pq}  + O(1/N^4)\,.
\eeq 
The eigenvalue problem \eqref{eigen_probl} is well defined as a consequence of the equality:
\beq
\bigg\{ \substack{ \#{\rm\ independent} \\[.1cm] {\rm entries\ of}\ \mathcal{A}\ \& \ \mathcal{M} }\bigg\}=
\bigg\{\substack{\#\ {\rm of\ } \langle \cO_{p_i}\cO_{p_j} K_{pq}\rangle \\[.1cm] +\,\ \#\ {\rm of}\ \eta_{pq} }\bigg\} \,.
\eeq
Let us comment on the structure of the matrices $\mathcal{A}$ and $\mathcal{M}$.
The SCPW expansion of disconnected free theory has the following compact expression:
\bea\nn
\cA^{\tau,l}_{a,b}= {\rm diag} \left( \mathcal{F}_{1+a+i+r,b-2r,r,a,t+a+r}\right)_{ \substack{1\leq i  \leq (t-1) \\ 0\leq r\leq \mu-1} }\,,\notag
\eea
where the function $\mathcal{F}$ is given by
\bea
&&
\cF_{p,\q,m,a,\var}= \tfrac{ p(p+\q)(1+\delta_{\q0}) (1+a)( 2m+2+\q+a)  (l+1)(l+2\var+2+\q)}{(p-1-m)!(p-2-m-a)!(p+m+\q)!(p+m+\q+1+a)! } \notag \\
&&
\rule{2cm}{0pt} 
\times 
\tfrac{ (m+1+\q)_{m+1} }{m!}\tfrac{ (m+2+a+\q)_{m+2+a} }{(m+1+a)!} \, \Pi_\var \, \Pi_{l+\var+1}\,, \nn \\
&&
\Pi_{\var}  \equiv \tfrac{ ((\var+\q)!)^2 }{(2\var+\q)! }  (\var+1-m)_m(\var+1+\q)_m (\var-m-a)_a \nn\\
&& 
\rule{1cm}{0pt} 
(\var+2+\q+m)_a  (\var+1-p)_{p-2-m-a}  \nn\\
&& 
\rule{1cm}{0pt}
(\var+3+\q+m+a)_{p-2-m-a}\,.
\eea
The SCPW of matrix elements in $\widehat{\mathcal{M}}_{\tau,l,a,b}$ has the form
\beq
\tfrac{\left(l+1+t+a+r+\frac{p_{43}-p_{21}}{2} \right)!\left(l+1+t+a+r+p_{43} \right)! }{\left(2(l+1+t+r+a)+p_{43}\right)!}\times \mathscr{P}_{d}(l)
\eeq
where $\mathscr{P}_d(l)$ is a polynomial in $l$ of degree $d={\rm min}( p_1+p_2, p_3+p_4)-(p_{43}-p_{21})-4 $, and $r$ labels $(p_3,p_4)$. 
We determine this polynomial case-by-case, and solve the eigenvalue problem following  \cite{Aprile:2017bgs,Aprile:2017xsp,Aprile:2017qoy}.  
We have verified that our conjecture (\ref{anomdims}) holds systematically in the $su(4)$ channels $[a,b,a]$ with $0\le a\leq 3 $, $0\leq b\leq 6$ up to twist $24$ for both even and odd spins.
In particular, we have been able to perform non-trivial tests on the pattern of residual degeneracies. It would be fascinating to understand how higher order corrections might lift the pattern of residual degeneracies observed at order $1/N^2$.

%%%%%%%%%%%%%%%%%%%%%%%%%%%%%%%%%%%%%%%%%%%%%%%%%%%%%%%
\section{IV. Casimir operators}
%%%%%%%%%%%%%%%%%%%%%%%%%%%%%%%%%%%%%%%%%%%%%%%%%%%%%%%

Quadratic and quartic conformal Casimir operators have played a useful role in understanding 
and simplifying the structure of correlators~\cite{Alday:2016njk,Alday:2017vkk,Aprile:2017qoy}. 
Here we extend the analysis of \cite{Aprile:2017qoy} to all $su(4)$ channels $[a,b,a]$ of any correlator $\langle p_1 p_2 p_3 p_4 \rangle$.
The quadratic and quartic Casimirs are given by~\cite{Alday:2016njk,Dolan:2003hv}
\begin{align}
	\mathcal{D}_2^{\rho_1,\rho_2} &= D^{\rho_1,\rho_2}_{+} + 2 \frac{x \bar{x}}{x-\bar{x}} \bigl((1-x)\partial_x - (1-\bar{x})\partial_{\bar{x}}\bigr)\,,\notag \\
	\mathcal{D}_4^{\rho_1,\rho_2} &= \biggl(\frac{x \bar{x}}{x-\bar{x}}\biggr)^2 D^{\rho_1,\rho_2}_{-} \biggl(\frac{x \bar{x}}{x-\bar{x}}\biggr)^{-2} D^{\rho_1,\rho_2}_{-}\,,
\label{conformalcasimirs}
\end{align}
where $D^{\rho_1,\rho_2}_{\pm}=D^{\rho_1,\rho_2} \pm \overline{D}{}^{\rho_1,\rho_2}$ and
\beq
	D^{\rho_1,\rho_2}=x^2 \partial_x(1-x)\partial_x-(\rho_1+\rho_2)x^2\partial_x-\rho_1\rho_2 x\,.
\eeq
The labels $\rho_i$ are given by $\rho_1=-\frac{1}{2}p_{12}$, $\rho_2=\frac{1}{2}p_{34}$. 
The eigenvalues of $\mathcal{D}_2$ and $\mathcal{D}_4$ on $\cB^{\left( 2+\frac{\tau}{2} |l \right)}$ are
\begin{align}
	\lambda_2(\tau,l) &=\tfrac{1}{2}(l(l+2)+(\tau+l)(\tau+l+4))\,,\notag \\
	\lambda_4(\tau,l) &=l(l+2)(\tau+l+1)(\tau+l+3)\,.
\end{align}
Consider the combination of Casimirs
\bea
\label{delta8}
	\Delta^{(8)}=-\tfrac{1}{8}&&
							\Big(\mathcal{D}_4^{\rho_1,\rho_2}-(\mathcal{D}_2^{\rho_1,\rho_2})^2+g^{a,b}_1 \mathcal{D}_2^{\rho_1,\rho_2}-g^{a,b}_2\Big) \nn\\
							\times&&\Big(\mathcal{D}_4^{\rho_1,\rho_2}-(\mathcal{D}_2^{\rho_1,\rho_2})^2+ g^{a,b}_3\mathcal{D}_2^{\rho_1,\rho_2}-g^{a,b}_4\Big),\qquad
\eea
with the coefficients $g^{a,b}_i$ given by
\begin{align}
	g_1^{a,b}&= (b+2 a)^2+6 (b+2 a)+6 ,\notag \\
	g_2^{a,b}&=\tfrac{1}{4}(b+2 a) (b+2 a+2) (b+2 a+4) (b+2 a+6),\notag \\
	g_3^{a,b}&=(b^2+2 b-2),\notag \\
	g_4^{a,b}&=\tfrac{1}{4}(b-2) b (b+2) (b+4).
\end{align}
The operator $\Delta^{(8)}$ has the property that its eigenvalue on the conformal blocks reproduces exactly the numerator of the anomalous dimensions given in equation~(\ref{anomdims}), i.e.
\beq
	\Delta^{(8)} \cB^{\left( 2+\frac{\tau}{2} |l \right)} = -2 M^{(4)}_{t} M^{(4)}_{t+l+1}\,\cB^{\left( 2+\frac{\tau}{2} |l \right)}.
\eeq

The operator $\Delta^{(8)}$ greatly simplifies the sums which compute the leading discontinuities of a correlator to any loop order.
In a large $N$ expansion we have
\beq
\mathcal{H}= \sum_{k\geq1} \frac{1}{N^{2k}} \sum_{r=0}^k \frac{1}{r!} (\log u)^r \, \sum_{m \leq n} \Upsilon_{nm} \mathcal{H}^{(k)}_{r,nm} \,.
\eeq
Then the leading discontinuity $\mathcal{H}^{(k)}_{k,nm}$ in an $su(4)$ channel with $a=n-m$ and $b=2m-p_{43}$ is given by
\beq
	\mathcal{H}^{(k)}_{k,nm} = \sum_{\tau,l,(q_1,q_2)} \left(\eta^{\tau,l,a,b}_{q_1q_2}\right)^k C_{q_1q_2}~\frac{\cB^{\left(2+\frac{\tau}{2} | l \right)}}{u^{2+\frac{p_{43}}{2}}},
\eeq
with $C_{q_1q_2}=\langle \mathcal{O}_{p_1} \mathcal{O}_{p_2} K_{q_1q_2} \rangle \langle \mathcal{O}_{p_3} \mathcal{O}_{p_4} K_{q_1q_2}\rangle$.
Since the numerator of the anomalous dimensions does not depend on $(q_1,q_2)$, 
we may pull out $(k-1)$ factors of $\Delta^{(8)}$ and remove $(k-1)$ powers of the numerator from the anomalous dimension. 
These reduced sums are considerably simpler.  
Indeed the resummed result for general $k$ is of a similar complexity as the $k=1$ case (the $\log u$ coefficient of the tree-level supergravity result). 
One can then recover the full leading discontinuity by applying $\Delta^{(8)}$ $(k-1)$ times to the resummed expression.

For concreteness, let us consider the simplest example: $p_i=2$, for which we have $\rho_1,\rho_2=0$ and the only $su(4)$ channel for long multiplets is the singlet $a,b=0$. 
The $(\log u)^2$ term of the $\langle 2222 \rangle$ correlator was computed at one loop  in~\cite{Aprile:2017bgs} (and recently reproduced using $\Delta^{(8)}$ in~\cite{Alday:2017vkk}). 
With the aid of $\Delta^{(8)}$ one can produce a closed formula for the highest transcendental weight part (weight $k$) of the leading $(\log u)^k$ discontinuity for any loop order: 
\begin{align}
	\mathcal{H}_{k}^{(k)}\Big|_{\rm top} &= \frac{1}{u^2} \bigl(\Delta^{(8)}\bigr)^{k-1} \Big[\frac{G_k(x,\bar{x}) - v^7 G(x',\bar{x}')}{(x-\xb)^7}\Bigr],\notag\\
G(x,\bar{x}) &= a_k(x,\bar{x}) \!\! \sum_{a_i =0,1} \! {\left[H_{a_10a_20\cdots1}(x)-(x\leftrightarrow\xb)\right]}.
\end{align}
Here $x'=\frac{x}{x-1}$ and $H_{c_1\cdots c_n}(x)$ are harmonic polylogarithms of weight $k$~\cite{Remiddi:1999ew}. 
Finally, the coefficient polynomial for the case $\langle2222\rangle$ is given by
\begin{align}
	a_k(x,\xb)& = - 2^{7-3 k} 3^{1-k}u^4\notag\\
	&\left[2^k (\hat{u}+v) \left(\hat{u}^2+8 \hat{u} v+v (v+6)\right)\right.\notag\\
	&-6 \left(\hat{u}^3+7 \hat{u}^2 v+3 \hat{u} v (v+2)-(v-4) v^2\right)\notag\\
	&\left.+ 5^{2-k} 2\left(\hat{u}^3-3 \hat{u} v^2+3 (\hat{u}+2) \hat{u} v+v^3\right)\right],
\end{align}
with $\hat{u}=u-1$. Similar results have been obtained for the correlators $\langle 2233 \rangle$, $\langle 2323 \rangle$, and $\langle 3333 \rangle$, 
for which the quantum numbers $(\rho_1,\rho_2)$ and $(a,b)$ of $\Delta^{(8)}$ are non-trivial. 

We believe that the results on the anomalous dimensions (\ref{anomdims}) together with the Casimir operators (\ref{delta8}) will aid in the construction of one-loop supergravity (i.e. order $1/N^4$) contributions to all correlators $\langle p_1 p_2 p_3 p_4 \rangle$. It would be fascinating to see if the methods described in \cite{Bargheer:2017nne} can be used to make contact with such supergravity loop corrections and the spectrum results described here.

\section*{Acknowledgements}
FA is supported by the ERC-STG grant 637844- HBQFTNCER.
JMD and HP acknowledge support from ERC Consolidator grant 648630 IQFT.
PH acknowledges support from STFC grant ST/P000371/1.

\vspace{-0.25cm}

%%%%%%%%%%%%%%%%%%%

\end{document}